\documentclass[a4wide,12pt]{article}
\usepackage[margin=1.2in]{geometry}
\usepackage{amsmath,amssymb,latexsym}
\usepackage{physics}

\usepackage{lmodern}

\usepackage{mathtools}
\usepackage[sorting=none, backend=bibtex]{biblatex}
\bibliography{refs}

\def \sec{\begin{section}}
\def \esec{\end{section}}

\renewcommand{\bar}{\overline}
\newcommand{\da}{\dagger}
\newcommand{\pp}{\prime}

\def \del {\delta}
\def \la {\lambda}

\def \om {\omega}

\def \ep {\epsilon}

\def \ka {\kappa}

\def \Oc {\mathcal{O}}

\def \pr {\partial}
\def \ra {\rightarrow}

\def \oh {\cfrac{1}{2}}

\def \beq { \begin{equation}}
\def \eeq {\end{equation}}

\def \OTOC {\text{OTOC}}
\def \id {\mathbf{1}}

\renewcommand\Re{\operatorname{Re}}

\newcommand\const{\operatorname{const}}

\def \l {\left}
\def \r {\right}

\def \bra {\langle}
\def \ket {\rangle}

\newcommand \bin[2] {
\begin{pmatrix}
#1 \\
#2
\end{pmatrix}}

\def \ee {\mathbb{E}}

\newcommand\AM[1]{\textcolor{blue}{{\it AM: #1}}}
\newcommand\JX[1]{\textcolor{purple}{{\it JX: #1}}}
\usepackage{authblk}
\begin{document}
\title{On scrambling, tomperature and superdiffusion in de Sitter space}

\author[1]{Alexey Milekhin\footnote{milekhin@caltech.edu}}
\author[2]{Jiuci Xu}

\affil[1]{Institute for Quantum Information and Matter, California Institute of Technology, Pasadena, CA 91125, USA}

\affil[2]{Department of Physics, University of California, Santa Barbara, CA 93106, USA}

\date{}
\maketitle

\begin{abstract}
    This paper investigates basic properties of the de Sitter static patch using simple two-point functions in the probe approximation. We find that de Sitter equilibrates in a superdiffusive manner, unlike most physical systems which equilibrate diffusively. We also examine the scrambling time.
In de Sitter, the two-point functions of free fields do not decay for sometime because quanta can reflect off the pole of the static patch. This suggests a minimum scrambling time of the order $\log(1/G_N)$, even for perturbations introduced on the stretched horizon, indicating fast scrambling inside de Sitter static patch.
We also discuss the interplay between thermodynamic temperature and inverse correlation time, sometimes called "tomperature."
    
\end{abstract}

\tableofcontents

\section{Introduction}
De Sitter spacetime is the simplest model of the expanding Universe.
Unfortunately, despite numerous efforts, holography for the de Sitter space remains mysterious.
This is mainly due to the fact that it is not clear where to put the holographic screen which hosts the dual quantum mechanical system. We refer to \cite{Banks:2001px,Banks:2003ta,Strominger:2001pn,Alishahiha:2004md,Gorbenko:2018oov,Coleman:2021nor,Anninos:2017hhn,Susskind:2021esx,Verlinde:2024znh} for the (possibly incomplete) list of the existing proposals. 
In this paper we would like to take a step back and revisit the general properties of de Sitter spacetime, without referring to holography.

\begin{figure}
    \centering
    \includegraphics{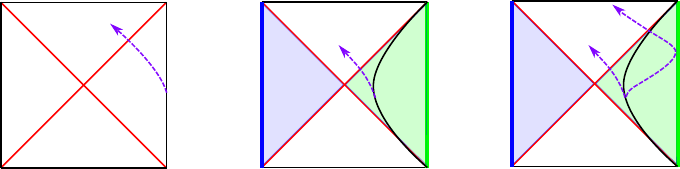}
    \caption{Illustrations of scrambling using Penrose diagrams of anti-de Sitter -- black hole (Left) and de Sitter (Center and Right). Left:  in anti-de Sitter the signal (purple) falls from the boundary inside the black hole horizon (red). 
    Once it is inside the stretched horizon it is scrambled.
    Center: naive picture in de Sitter. The signal starts at the stretched horizon (black) and quickly falls inside the cosmological horizon.
    Right: our proposal for how free bulk fields scramble in de Sitter. A part of the initial perturbation falls towards the pole first. This takes a long time (of order $2 R_{dS} \log(1/G_N)$), this is why the scrambling takes at least this time.}
    \label{fig:scramble}
\end{figure}

We will mostly concentrate on the case when de Sitter is almost empty. One immediate problem with de Sitter is the absence of naturally defined time, because it is a closed universe. In this paper we will follow the approach of \cite{Chandrasekaran:2022cip} and postulate the presence of an observer. For simplicity we will assume that they do have a Hamiltonian and hence a clock, but otherwise interact weakly with the de Sitter matter fields. 
The observer can explore a part of de Sitter which is called static patch.
 This setup is intended to mimic our own experience in the Universe.
The static patch has the following metric in $d-$dimensions:
\beq
\label{eq:static_metric}
\frac{ds^2}{R_{dS}^2} = -(1-r^2) dt^2 + \frac{dr^2}{1-r^2} + r^2 d\Omega_{d-2}^2.
\eeq
In the semi-classical states, where the observer's clock has a definite reading, 
 the observer's time is simply proportional to the proper time at the pole $r=0$, that is, the coordinate $t$.
In this paper we will measure everything in the units of de Sitter radius. Static patch is observer dependent and only covers a part of the whole de Sitter. The Penrose diagram of the full de Sitter is illustrated by Figure \ref{fig:scramble} (Center and Right), the static patch is shaded by green.

Unlike graviton and matter higher-point correlation functions, which may include the backreaction on the metric and depend on how we introduce the holographic screen, simple two-point functions in the probe approximation are unambiguous and easy to compute. They probe the simplest possible physical process, namely how an initial perturbation spreads. Our goal is to extract as much physical information from them as possible. As we will see, they do contain a lot of interesting physics.

One characteristic feature of the static patch (\ref{eq:static_metric}) is the presence of a horizon, the so-called cosmological horizon, at $r=1$. 
Matter or graviton perturbations fall under the horizon and thus the spectrum of linearized perturbations has complex frequencies. These are the famous quasi-normal modes. 
First, we will discuss how fast an initial perturbation spreads around de Sitter.
  In Section \ref{sec:superdiff} we will argue that an initial perturbation spreads around de Sitter superdiffusively: the displacement on the transverse $S^{d-2}$ ($d\Omega_{d-2}^2$ in the line element (\ref{eq:static_metric})) is proportional to time \footnote{More generally, superdiffusion/subdiffusion means any propagation faster/slower than diffusion.}. More precisely, the spreading is controlled by the following equation:
  \beq
  \pr_t \rho \approx -\sqrt{-\Delta_S} \rho,
  \eeq
  where $\Delta_S$ is the sphere Laplacian.
This is surprising, because most physical systems (including black holes in anti-de Sitter) equilibrate diffusively:
\beq
\pr_t \rho = D \Delta \rho,
\eeq
where the displacement  is proportional to the square-root of time. This behavior persists even if de Sitter is not empty, but has a black hole inside of it. Whatever is the quantum mechanical dual to de Sitter static patch, it has to reproduce this behavior.

In addition to spacial spreading, correlations also decay in time.
One of the most widely accepted conjectures about de Sitter is that empty static patch should be described by a maximally-mixed density matrix \cite{Bousso:2002fq,Dong:2018cuv}.
That is, the corresponding Boltzmann temperature is infinite.
Such states rarely
arise
\footnote{With the exception of random quantum circuits.}
in physical systems.
In Section \ref{sec:infT} we will offer a few comments on the interplay of temperature and the decay rate, for which ref. \cite{Lin:2022nss} coined the term "tomperature".  
 We would like to point out that the maximally mixed state has one very characteristic feature. Namely, for a local Hamiltonian system one expects the correlations to decay the fastest at infinite temperature. Specifically, assume that the decay in time is exponential: 
\beq 
\Tr (\rho \Oc(t) \Oc(0)) \propto e^{-\kappa t}.
\eeq
Then one should expect that $\kappa$ ("tomperature" in the terminology of \cite{Lin:2022nss}) is maximal for infinite thermodynamic temperature, when $\rho \propto 1$. 
We will not attempt to prove this statement rigorously, instead we will illustrate this statement with various physical Hamiltonian examples.
Ideally, we would like to see the counterpart of this in de Sitter, but we had only limited success with this.

The last section is dedicated to the scrambling time, the time when the initial perturbation completely delocalizes. It has been argued recently \cite{Susskind:2021esx} that the scrambling time in de Sitter can be hyperfast, namely of order 1 in the units of the de Sitter radius if the initial perturbation is introduced near the cosmological horizon - Figure \ref{fig:scramble} (Center).  For fields which are confined to the horizon this is a reasonable expectation. But in holography we are usually interested in free bulk fields. What happens to them?
We will point out that for free bulk fields hyperfast scrambling is unlikely. An honest computation of the scrambling time requires a careful evaluation of the out-of-time ordered correlation (OTOC) function and we will not attempt it here.
Our argument is based on the fact that for free bulk fields two-point function does not decay for a long time, time of the order $2 R_{dS} \log(1/G_N)$. This originates from the fact that a field introduced near the cosmological horizon can fall towards the pole first, reflect back, and only after that fall inside the cosmological horizon - Figure \ref{fig:scramble} (Right). This takes time of order $\log(1/G_N)$ because one needs to go away from the stretched horizon and then back. Hence the information is still localized in the bulk prior to this time and can be accessed from the stretched horizon using a simple correlation function.
Whereas the phenomena of scrambling is associated with complete delocalization, when initial perturbation cannot be accessed using any simple correlation functions. Hence scrambling cannot happen prior to the time when the two-point function decays, which is $2 R_{dS} \log(1/G_N)$. 
To back up this statement, we will provide a few lower bounds on OTOC in qubit systems using two-point functions. This will quantify the intuition that if two-point function is not small, OTOC cannot be small.

Various technical details are dedicated to the Appendices. 
\section{Superdiffusion}
\label{sec:superdiff}
Suppose we want to study perturbations, either matter or graviton, inside the Sitter static patch (\ref{eq:static_metric}).
Imposing the regularity inside ($r<1$) and the in-going boundary conditions at the cosmological horizon ($r=1$) leads to a series of quasi-normal modes. There is a discrete series of them for each angular momentum $l$ associated with $S^{d-2}$.
 For example, for a massive field of mass $m^2 = \Delta(d-1-\Delta)$ minimally coupled to gravity the answer \footnote{There is also the second branch
 $\omega_k = -i ( l + \Delta + 2k + d-1)\kappa_c$.
 }\cite{Lopez-Ortega:2006aal} is 
\beq
\label{eq:ds_qnm}
\omega_k = -i ( l + \Delta + 2k )\kappa_c, \ k=0,1,2,\dots,
\eeq
where $\kappa_c = 1 $ is the surface gravity of the cosmological horizon.

Before extracting some physical lessons from it, let us recall the quasi-normal frequencies for a planar black hole in AdS \cite{Policastro:2002se}. In this case there is a diffusive branch with the frequencies parametrized by the planar momentum $p$ along the horizon:
\beq
\label{eq:diff}
\omega = -i D p^2, \ p \ra 0, \ D=\const.
\eeq
This leads to the diffusion along the horizon: $\pr_t \rho = D \Delta \rho$, which also says that systems dual to anti-de Sitter equilibrate diffusively. Diffusion is ubiquitous in our everyday experience and it also controls more special systems, such as $\mathcal{N}=4$ super Yang--Mills and Sachdev--Ye--Kitaev chains \cite{Gu2017Local,Song2017Strongly,Milekhin:2021cou}.
Returning to de Sitter and the equation (\ref{eq:ds_qnm}), we see that instead of diffusion we get the so-called "superdiffusion" on the sphere, because $l$ is proportional to the square-root of the sphere Laplacian $\Delta_S$:
\beq
\label{eq:sdiff}
\pr_t \rho \approx - \sqrt{-\Delta_S} \rho.
\eeq
This equation becomes valid when the quasi-normal mode ringdown starts, which should be after a few light-crossing times in de Sitter.
This superdiffusive spread is much faster than the diffusive one, as the distances are proportional \footnote{ Sometimes such behavior is called ballistic because the spread is proportional to time. However, we would like to abstain from this terminology since eq. (\ref{eq:sdiff}) is dissipative because the Fourier components decay.
A typical example of ballistic propagation is a wave-equation $(\pr_t^2 - \pr^2_x) \rho = 0 $ which does not involve the decay of the Fourier components.
} to time. For $2+1$ de Sitter it is exactly the square-root of the Laplacian. For higher-dimensions it becomes the square-root\footnote{For $(d-2)$-dimensional sphere the eigenvalues of the Laplacian are $-l(l+d-3)$.} only for large $l$, which correspond to perturbations localized on the sphere.
Unlike diffusion which is ubiquitous, appearance of superdiffusion is rare.
Examples include integrable or near-integrable chains \cite{Bulchandani:2021rzz} and systems with long-range interactions \cite{Deng:2022puf,Juh_sz_2008,Moyano_2006}.
Whatever is holographically dual to de Sitter, it must reproduce this superdiffusive feature. What we said is true not only for classical perturbations, but for weakly-interacting quantum fields inside de Sitter as well, because for free fields $\phi$ the retarded Green function is determined by the commutator $[\phi(x_1) , \phi(x_2) ]$ which is c-number and it depends only on the form of the mode functions and not the quantum state.

This conclusion hinges on the linear $l$-dependence of the quasi-normal frequency and we would like to claim that it is robust feature of de Sitter.
It is a feature for all tensor/vector/scalar perturbations, the answer for them differs only by an $l$-independent constant.
Also it pertains to de Sitter-Schwarzschild spacetimes: we analyze $2+1$-dimensional case in the Appendix \ref{app:ds_bh} and for higher-dimensions we refer to \cite{Brady:1996za, Brady:1999wd, Molina:2003dc, Konoplya:2022xid}.
Unfortunately, these computations have only been done numerically. In Appendix \ref{app:deform} we analyse the case when we still have an undeformed even-dimensional empty de Sitter but we modify the scalar field boundary condition at the pole. Interestingly, we find that for a massless field the lowest quasi-normal mode $\omega = - i l k_c$ exists for all boundary conditions. For a massive scalar the expression changes, but the linear dependence on $l$ remains, at least for large $l$. It would be interesting to perform a similar analysis when the spherical symmetry is broken.

Are there any other spacetimes for which we expect superdiffusive behavior? Naively, if we put the holographic screen very close to the horizon we can expect the equilibration to happen very fast. This scenario can arise if we have a very big black hole in anti-de Sitter, or we $T\bar{T}$-flow the dual field theory close to the black hole horizon. 
In Appendix \ref{app:diff} we analyze this case, but surprisingly only find normal diffusion.
Note that this setup is different from the de Sitter static patch case which we primarily addressed in this paper, where there is also the portion of spacetime between the screen and the pole.

 It would be nice to understand the superdiffusion in de Sitter on a more intuitive level.  We conjecture that it is related to the transverse sphere capping off at the pole. In the Appendices \ref{app:ds_bh} and \ref{app:toy} we study three metrics which are slightly deformed compared to $2+1$ de Sitter static patch (\ref{eq:static_metric}):
 
 de Sitter black hole 
\beq
ds_{BH}^2 = -(1-2M -r^2)dt^2 + \frac{dr^2}{1-2M-r^2}+r^2 d\varphi^2,
\eeq
and two toy ones which we refer to as geometries A and B in the following:
\beq
ds_A^2 = -(1-r)dt^2 + \frac{dr^2}{1-r} + r^2 d\varphi^2,
\eeq
\beq
ds_B^2 = -(1-r^2)dt^2 + \frac{dr^2}{1-r^2} + d\varphi^2.
\eeq
Three-dimensional de Sitter black hole and geometry A still contain a capping off sphere, but the time-radial geometry is modified \footnote{Strictly speaking, geometry $A$ has a singularity at $r=0$ but we ignore that.}. For geometry B instead of a sphere we have a cylinder which does not cap. For the dS-black hole and geometry A we find superdiffusion for scalar fields, whereas for geometry B we find a normal diffusion.

Let us close this Subsection with a comment about the microscopic origin of (super-) diffusion. The standard diffusion equation (\ref{eq:diff}) can be obtained from an ensemble of particles undergoing Brownian motion, where for each timestep $\Delta t$ the displacement $\Delta x$ is determined by the short-range Gaussian distribution $e^{-(\Delta x)^2/\Delta t}$. 
Superdiffusion (\ref{eq:sdiff}) can be obtained by a similar manner, but now the particles are undergoing the so-called L\'evy flight: the displacement $\Delta x$ is determined by the long-range Cauchy distribution $1/(1+(\Delta x/\Delta t)^2)$.

\subsection{A comment on temperature and
tomperature}
\label{sec:infT}
As we mentioned in the Introduction, one of them most prominent features of de Sitter static patch is its infinite temperature.
In this Section we would like offer some comments on this result by studying the behavior of correlation functions, specifically the "tomperature", also known as inverse correlation time. However, unlike the rest of the paper, in this Section we will not reach any sharp conclusions.

Let us first present some intuitive physical picture and then we will talk about the caveats.
For a moment, let us put aside de Sitter and let us consider an infinite temperature state.
Intuitively, infinite temperature state is supposed to be the most disorganized one, where the correlation functions decay the fastest. For example, assume the exponential decay at late times:
\beq
\label{eq:infT}
\Tr( \Oc(t) \Oc(0) ) \sim e^{-\kappa t}.
\eeq
Inverse correlation time $\kappa$ is what \cite{Lin:2022nss} dubbed "tomperature".
Then in other states we can expect the decay constant to be smaller:
\beq
\label{eq:notinfT}
\Tr(\rho \Oc(t) \Oc(0) ) \sim e^{-(\kappa-\delta \kappa) t}, \ \delta \kappa > 0.
\eeq
This is true for any 1+1 conformal field theory \footnote{Correlators at finite temperature and infinite volume are fixed by the conformal symmetry to be $\bra \Oc_\Delta(t,x) \Oc_\Delta(0,0) \ket \sim \sinh(\pi T (t \pm x))^{2\Delta}$. The greater the temperature the faster they decay.}, large-$p$ Sachdev--Ye--Kitaev (SYK) \cite{SachdevYe,KitaevTalks,ms} model \footnote{If the SYK coupling is $\mathcal{J}$, then the decay rate $\kappa$ at inverse temperature $\beta$ is determined by \cite{ms}:
$ \mathcal{J} \cos \frac{\beta \kappa}{2}= \kappa$. Decay rate $\kappa$ monotonically increases with increasing temperature. For $\beta=0$, $\kappa=\mathcal{J}.$
}, Brownian SYK model \cite{Milekhin:2023bjv}, XX spin chain \cite{Stolze_1995}. This statement seems intuitive, but it would be instructive to put it on firmer footing.
As a side note, in the condensed matter systems it is expected that local dissipation rate $\kappa$ obeys a 'Planckian' bound \cite{Lucas:2018wsc,Almheiri:2019jqq,Hartnoll:2021ydi}, $\kappa \le T/\hbar$.
Note that we are talking about the decay in time, rather than space. 
Spacial correlation length has very different properties.
For a free massive field, the spatially separated correlator decays exponentially, whereas for time-like separation it oscillates. 

Is it true that in empty de Sitter correlations decay the fastest? For example, compared to de Sitter--black hole. Yes and no.
In empty de Sitter a generic perturbation of a scalar field of mass $\Delta$ decays exponentially at late times:
\beq
\label{eq:decay}
\phi \sim e^{- \Delta \kappa_c t},
\eeq
where $\kappa_c$ is the surface gravity of the cosmological horizon and $t$ is the standard static patch time. This is the consequence of the quasi-normal modes (\ref{eq:ds_qnm}).
Exponential decay holds even in the presence of backreaction on the geometry \cite{Brady:1996za}.
It turns out that eq. (\ref{eq:decay}) remains valid
for non-empty de Sitter, for example for de Sitter -- Schwarzschild when the black hole is not too big \cite{Brady:1999wd, Molina:2003dc,Konoplya:2022xid,Jansen:2017oag}, where $\kappa_c$ is still the surface gravity of the cosmological horizon. Crucially, $\kappa_c$ is always \textit{smaller} than in the empty de Sitter. Exactly as we would have expected if we deformed away from an infinite temperature state!  

Now let us mention the caveats. 
So far we have been using the Schwarzschild time $t$. Obviously, this is not diff-invariant. Our point is that for quantum systems we expect the correlation time to be minimal at maximal temperature and holography for de Sitter space has to somehow meet this expectation.
 We can sensibly compare the decay rates in different spacetimes if we introduce an observer. In Appendix \ref{app:dressing} we discuss
different ways how it can be done. It turns out that sometimes the decay rate can become larger if the spacetime is deformed away from the empty de Sitter due to the observer's warp factor.
Another caveat concerns the matter boundary conditions. In Appendix \ref{app:deform}
we analyze the case when we slightly change the boundary conditions at the pole. We observe that the correction to the quasi-normal frequency can have either sign.
Thus it is not strictly true that in empty de Sitter correlations decay the fastest.


\section{What is the scrambling time in de Sitter?}
\label{sec:scrambling}

Let us start from recalling the scrambling time computation in anti-de Sitter. We insert a perturbation at the boundary which then falls under the black hole horizon - Figure \ref{fig:scramble} (Left). 
The boundary (Schwarzschild) time it takes to reach the stretched horizon is $\beta \log(1/G_N)$. Basically all of this time comes from the infinite time dilation near the horizon. Before this time the signal is still localized in the bulk, although one might need a non-local boundary observable to probe it.

\begin{figure}
    \centering
    \includegraphics{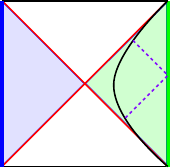}
    \caption{A null geodesic (purple) connecting two points on the stretched horizon (black), resulting in a very large correlation.}
    \label{fig:bounce}
\end{figure}

Now consider de Sitter.
If the holographic screen is away from the cosmological horizon then the picture is very similar to anti-de Sitter and we would expect fast scrambling time of $R_{dS} \log(1/G_N)$.
What if we put the holographic screen right on the stretched horizon? 
Naively, the signal falls right inside the horizon, so the scrambling time should be of order 1 - Figure \ref{fig:scramble} (Center). This picture is perfectly reasonable if the field is somehow confined to the horizon \cite{Susskind:2021esx} or the spacetime to the right of the stretched horizon is absent altogether. We would like to understand what happens if we deal with a free bulk field and the complete spacetime.
 In this case, an initial perturbation will go \textit{inside} the static patch as well - Figure \ref{fig:scramble} (Right). It will be present there for some time: first it will go towards the pole, reflect back and only after that fall inside the horizon. So the signal remain localized in the bulk for a long time, of order $2 R_{dS} \log(1/G_N)$, this is the time needed to go to the pole and back. Notice that this time is actually twice as big compared to the (estimation of) scrambling time when the holographic screen is near the pole. Moreover, on its way back, when it is crossing the stretched horizon, it will become highly localized again, resulting in a very big two-point function.  In fact, due to the transverse sphere, there is a family of different null geodesics - Figure \ref{fig:bounce}. Because of them two-point functions at different angular separations will be large prior to $t \sim 2 R_{dS}\log(1/G_N)$. So the presence of the original perturbation can be probed in a very simple way by computing a two-point function of local operators.

What is the implication of this on scrambling? 
Scrambling is delocalization of information when the initial perturbation cannot be probed using simple correlation functions. Here we see that in de Sitter prior to time of order $\log(1/G_N)$ the initial perturbation is still very much localized. 
We conclude that scrambling time is at least $\log(1/G_N)$. So we should expect the usual fast scrambling instead of hyperfast scrambling.

Let us do a simple bulk two-point function computation in more detail.
The two-point function of a massive scalar in Bunch--Davies (Euclidean) vacuum of 3-dimensional de Sitter is 
\beq
\bra \phi \phi \ket = G = \phantom{m}_2 F_1 \l( \Delta_+,\Delta_-,\frac{3}{2}; \frac{1+P}{2} \r),
\eeq
where $P = \cos D$, $D$ being the geodesic distance and $\Delta_{\pm}$ determine the mass of the scalar by $m^2 = \Delta(2-\Delta)$. For points inserted at the stretched horizon $r=r_{sh} \approx 1 - G_N$, at times $0,t$ and angular reparation $\theta$, it reads as (here $R_{dS}=1$)
\beq
P = 1 + 2 (1-r_{sh}^2) \sinh^2(t/2) -2r_{sh}^2 \sin^2 \frac{\theta}{2}.
\eeq

\begin{figure}
    \centering
    \minipage{0.45\textwidth}
    \includegraphics{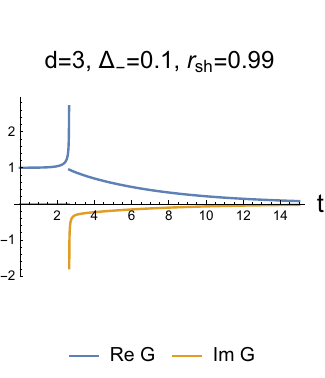}
    \endminipage
     \minipage{0.45\textwidth}
    \includegraphics{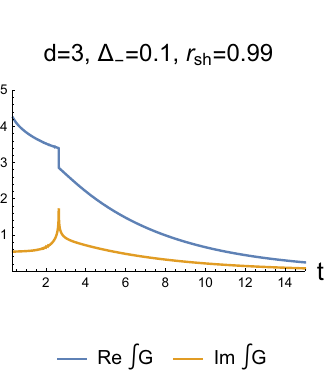}
    \endminipage
    \caption{Behavior of the two-point function of a massive scalar field in 3-dimensional de Sitter. Left: operators localized on the transverse sphere. Right: operators uniformly smeared over the sphere.}
    \label{fig:gs}
\end{figure}

For operators inserted at a point on the $S^1$ the two-point function blows up for separations of order $-\log(1-r_{sh}) \sim \log(1/G_N)$, suggesting that scrambling cannot happen prior to that time, because this blow up indicates that the originally inserted information can be recovered using the said two-point function. In fact, while $2(1-r_{sh}^2) \sinh(t/2)^2$ is less than 1, different two-point functions blow-up depending on $\theta$. We plot $G$ for some sample values at Figure \ref{fig:gs} (Left).

Instead of studying the correlations of signals inserted at a specific location on the transverse sphere, we can also smear the signal over the sphere. For that we need to look at the smeared two-point function 
\beq
\int d\theta_1 d\theta_2 \langle \phi(t_1,\theta_1,r_{sh}) \phi(t_2,\theta_2,r_{sh}) \rangle.
\eeq
It is illustrated by the Figure \ref{fig:gs} (Right)
For the smeared Green function is picture is a bit less clear because the initial value of $G$ at $t=0$ is infinity (not shown). The presence of the time-like geodesics now results in finite jump at finite time - Figure \ref{fig:gs} (Right).
However, given the presence of the structure at $t=\log(1-r_{sh})$ (the glitch in the terminology of \cite{Rahman:2022jsf}) and the fact that the two-point function still continues to decay past that point, we conjecture that scrambling again does not happen prior to $-\log(1-r_{sh}) \approx \log(1/G_N)$.

A more direct way to probe scrambling is the OTOC. As we said, we will not attempt to perform the computation here, but we would like to comment that at infinite temperature two-point functions can be used to bound the OTOC from below in an effective way. This is in accord with our statement that scrambling (OTOC is small) can happen  only when two-point functions are sufficiently small.
In Appendix \ref{app:otocbound} we rigorously proof that for any operators $\Psi_1, \Psi_2$, which are a product of Majorana operators (or spins), the following inequality holds at \textit{infinite temperature}:
\beq
|\text{OTOC}| = |\bra \Psi_1(t) \Psi_2(0) \Psi_1(t) \Psi_2(0) \ket| \ge 2\bra \Psi_1(t) \Psi_1(0) \ket^2 - 1.
\eeq
The idea behind the proof is that small OTOC requires the Heisenberg-evolved operator $\Psi_1(t)$ to be widespread. Whereas large auto-correlation function says that the operator is still mostly localized.
This bound is very general, it does not rely on the number of spacial dimensions or any locality in the Hamiltonian. The only input is that the Hamiltonian is fermion-even.
This bound holds even for integrable systems, in fact it is saturated for a harmonic oscillator.

Unfortunately, it does not say that much, only that the fermion two-point function must decrease from $1$ to $1/\sqrt{2}$ for the OTOC to decay. In Appendices \ref{app:spinOTOC} and  \ref{app:typical} we provide analytical evidence that for chaotic systems at early times (before the two-point function decays) one can expect a stronger bound:
\beq
|\bra \Oc_A(0) \Oc_D(t) \Oc_A(0) \Oc_D(t) \ket| \ge |\bra \Oc_A(t) \Oc_A(0) \ket|^2,
\eeq
where $\Oc_A, \Oc_D$ are distinct qubit operators, not necessarily local. This bound says that OTOC cannot be small unless two-point function is small.
In Appendix \ref{app:otocnum}
we study this bound in numerical examples. We find that it holds for different chaotic systems, but can be violated for integrable or near-integrable systems.

\section{Conclusion}
In this paper we addressed three features of de Sitter: superdiffusive spread of perturbations, fast scrambling time $\sim \log(1/G_N)$ for free bulk fields and finally the decay rate of the correlations.
These features are intrinsic to de Sitter and our arguments did not depend on the location of the holographic screen.

Superdiffusion suggests that the quantum mechanical theory describing the static patch must be rather special. An "easy" way to get a superdiffusion in a system is to introduce a long-range hopping.
We conjectured that the reason why de Sitter is superdiffusive is because the geometry is capping off at the pole. Our logic is based on a naive picture of perturbations spreading very fast on the sphere near the pole because the sphere is very small there. We analytically analyzed quasi-normal frequencies in several "deformed de Sitter" geometries in the Appendix \ref{app:qnm} and our results are consistent with this conjecture. Of course, it would be interesting to provide more general arguments.

Our argument that scrambling of free bulk fields cannot happen prior to $\log(1/G_N)$ was indirect: we studied the behavior of two-point function which does not decay prior to that time. Then we argued that scrambling cannot happen before local correlations decay. It would be instructive to perform an honest bulk computation of the OTOC. Similar calculation has been carried out in the literature before \cite{Aalsma:2020aib,Geng:2020kxh}, but they placed the initial perturbation near the pole rather than the stretched horizon. They found fast scrambling behavior in accord with naive expectations. Presumably, when a perturbation is introduced near the stretched horizon, the pole-reflection effect (Figure \ref{fig:scramble} (Right) and Figure \ref{fig:bounce}) becomes important for the shock-waves.

Finally, in Section \ref{sec:infT} we speculated that the infinite temperature of de Sitter is reflected in the decay time of correlation functions. In many physical examples with local Hamiltonians the correlation functions decay the fastest at infinite temperature. Similar behavior can be observed for the cosmological horizon surface gravity, which often controls the quasi-normal decay rate. Unfortunately, for closed spacetimes it is not a well-posed question to compare the decay rates for different spacetimes. We found that the decay rate is highly observer-dependent. Also, for matter fields perturbations the cosmological horizon surface gravity is not the only thing which determines the decay rate as we saw in Appendix \ref{app:deform}.

Another comment which is unrelated to the questions we studied in this paper concerns the so-called scar states. This is class of states in quantum many-body systems which have finite energy density, but are not thermal. By that we mean that they have low entanglement entropy and often support eternal oscillations in the correlation functions, we refer to \cite{serbyn2021quantum}
 for a review. Such behavior is atypical for chaotic systems, but it has been even observed experimentally \cite{Bernien_2017,turner2018weak}.
A recent paper \cite{Milekhin:2023was} argued that all holographic systems dual to anti de-Sitter gravity have scar states. The reason is simple: instead of a black hole in the center of anti-de Sitter one can put a boson star \cite{Buchel:2013uba,Buchel:2014dba} or a gravitational wave \cite{Horowitz:2014hja,Liska:2022vrd,Caputa:2022zsr}. The stress-energy tensor at the boundary will be non-zero, but the entanglement entropy is low. We expect that systems dual to de Sitter have scars too because de Sitter supports boson star solutions as well (e.g. \cite{Kumar:2014kna,Kumar:2017zms}). Another approach to holographic scars is based on periodic orbits around black holes \cite{Dodelson:2022eiz}.

Let us mention a few purely quantum-mechanical questions which would be interesting to study separately from gravity.
First of all, we did not prove rigorously that the correlations decay the fastest at infinite temperature. This is just an expectation and we provided various examples of this phenomena in a diverse range of systems. Also, we numerically observed that in chaotic qubit/spin systems
the OTOC at early times can be bounded from below by the square of two-point function $G^2$, for both local- and non-local operators. In Appendix \ref{app:otocbound} we gave an argument for this bound assuming that the operator weight distribution is monotonic.
This is a refinement of the bound $|\text{OTOC}| \ge 2G^2-1$. 
It would be interesting to study both of these question more rigorously,  especially exploring whether they can be elucidated within the algebraic formalism recently developed in \cite{xu2024von}.

\section*{Acknowledgement}
We are grateful to Ying~Zhao for collaboration at the early stages.
We would like to thank 
Xi~Dong, Hao~Geng,
Daniel~Harlow,
Alexei~Kitaev, Juan~Maldacena, Donald~Marolf, John~Preskill, 
Tommy~Schuster for comments and discussions. We are thankful to Eugenia~Colafranceschi,
Gary~Horowitz,
Adel~Rahman and Leonard~Susskind for the comments on the manuscript.

AM acknowledges funding provided by the Simons Foundation, the DOE QuantISED program (DE-SC0018407), and the Air Force Office of Scientific Research (FA9550-19-1-0360). The Institute for Quantum Information and Matter is an NSF Physics Frontiers Center. AM was also supported by the Simons Foundation under grant 376205. J.X. was on the MURI grant and was supported in part
by the U.S. Department of Energy under Grant No. 
DE-SC0011702. This material is
based upon work supported by the Air Force Office of Scientific Research under award
number FA9550-19-1-0360.

\appendix

\section{Quasi-normal modes}
\label{app:qnm}
\subsection{Empty de Sitter with alternative boundary conditions}
\label{app:deform}
Let us re-derive the quasi-normal modes for de Sitter.
Following \cite{Lopez-Ortega:2006aal} we introduce the variable $y=r^2$. Then the radial part of the wave equation in de Sitter static patch can be written as
\beq
y(1-y) \phi'' - \frac{1}{2}((d+1)y-(d-1)) \phi' + \frac{1}{4} \l( \frac{\omega^2}{1-y} - \frac{l(l+d-3)}{y}-m^2 \r) \phi = 0.
\eeq
The solutions are
\beq
\phi_1 = y^{l/2} (1-y)^{-i \omega/2} \phantom{m}_2 F_1 \l( \frac{l+\Delta-i\omega}{2},\frac{d+l-1-\Delta-i\omega}{2},l+\frac{d-1}{2},y \r),
\eeq

\beq
\phi_2 = y^{3/2-l/2-d/2} (1-y)^{-i \omega/2} \phantom{m}_2 F_1 \l( \frac{2-l-\Delta-i\omega}{2},\frac{3-d-l+\Delta-i\omega}{2},\frac{5-d-2l}{2},y \r).
\eeq
$\phi_2$ only has this form if $d$ is even. Otherwise $(5-d-2l)/2$ becomes a negative integer and one has to use an alternative form of the formula for $\phi_2$.

Obviously, only $\phi_1$ is regular at the pole $y=r^2=0$. To enforce the in-going boundary conditions at the horizon it is natural to perform the standard $y \ra 1-y$ transformation of the hypergeometric function and require the absence of $(1-y)^{+i \omega/2}$. 
The transformation is
\begin{align}
\label{eq:z_transform}
\phantom{m}_2 F_1(a,b,c,y) = 
\frac{\Gamma(c)\Gamma(c-a-b)}{\Gamma(c-a)\Gamma(c-b)}\phantom{m}_2 F_1(a,b,a+b-c+1,1-y) + 
\nonumber \\ + \frac{\Gamma(c)\Gamma(a+b-c)}{\Gamma(a) \Gamma(b)} (1-y)^{c-a-b}
\phantom{m}_2 F_1(c-a,c-b,1+c-a-b,1-y).
\end{align}
We can ensure that one of the terms vanishes if a Gamma function in the denominator has a pole.
Doing this for $\phi_1$ leads to two branches of quasi-normal modes:
\beq
\label{eq:low_branch}
\omega = -i (\Delta + l + 2k),
\eeq
\beq\label{eq:branch2}
\omega = -i (-\Delta + l + d -1 + 2k), \ k=0,1,2,\dots.
\eeq
Suppose we somehow modify the boundary conditions at the pole. Now we need to make sure that a linear combination  $\phi_1 + \epsilon \phi_2$ (for some phenomenological $\epsilon$) has in-going boundary conditions. This will be true if the following combination vanishes:
\beq
\label{eq:vcombo}
\frac{\Gamma(l+\frac{d-1}{2})}{\Gamma(\frac{l-i \omega+\Delta}{2}) \Gamma(\frac{-1+d+l-i \omega-\Delta}{2})} + \epsilon \frac{\Gamma(-l+\frac{5-d}{2})}{\Gamma(\frac{2-l-i \omega-\Delta}{2}) \Gamma(\frac{3-d-l-i \omega+\Delta}{2})}.
\eeq
In general, it can be a complicated equation because $\ep$ can depend on $l$ and $\omega$.
We can find the correction to the lowest mode $-i (\Delta + l)$ perturbatively, assuming that $\epsilon$ is small. The new value is given by
\beq
\omega = -i (\Delta + l + \delta),
\eeq
\beq
\delta = - \epsilon \frac{2 \Gamma(\frac{-1+d-2\Delta}{2})}{\Gamma(l+ \frac{d-3}{2}) \Gamma(1-l-\Delta)}.
\eeq
For large $l$ this correction goes as $l^{\Delta-(d-3)/2}$. 
If $\ep$ grows slower with $l$ then we again have
superdiffusion. Interestingly, for massless case the second term in eq. (\ref{eq:vcombo}) is actually equal to zero at $\omega = - i l$, so this mode is not shifted.

As an example, consider Dirichlet boundary condition $\phi_1(r_c) + \ep \phi_2(r_c)=0$. 
Parameter $\epsilon$ is determined by
\beq
\epsilon = -r_c^{2l+d-3}\frac{\phantom{m}_2 F_1 \l( \frac{l+\Delta-i\omega}{2},\frac{d+l-1-\Delta-i\omega}{2},l+\frac{d-1}{2}, r_c^2 \r)}{\phantom{m}_2 F_1 \l( \frac{2-l-\Delta-i\omega}{2},\frac{3-d-l+\Delta-i\omega}{2},\frac{5-d-2l}{2}, r_c^2 \r)}.
\eeq
Let us assume that $\epsilon$ and the correction to the frequency is small. Then we can plug in $\omega = -i(\Delta+l)$ inside $\ep$, resulting in:
\beq
\ep =  \frac{2}{(2l+d-3)B(r_c^2,\frac{3-d}{2}-l,l+\Delta)},
\eeq
where $B$ is incomplete Beta-function:
\beq
B(y,a,b) = \int_0^y dx \ t^{a-1} (1-t)^{b-1}.
\eeq
For large $l$, if $r_c^2<1/2$ then this Beta-function is large $\sim r_c^{-2l}$. Otherwise it is small $\sim 1/l$, making $\epsilon$ a finite constant. Hence if $r_c^2 < 1/2$ (which is, in fact, exactly half-way between the pole and the horizon), then superdiffusion persists.

\subsection{Diffusion near the horizon}
\label{app:diff}
Suppose we put the holographic screen near the horizon and ask what happens with the transport in the dual theory. 
The metric can be approximated as
\beq
ds^2 = -T(1-r)dt^2 + 
\frac{dr^2}{T(1-r)} + ds_M^2,
\eeq
where $ds^2_M$ denotes line element on some spacial manifold $M$. We will denote the eigenvalues of the corresponding Laplacian by $-p^2$. $T$ is the surface gravity of the horizon. Spacetime only exists between $r=1$ and $r_* < 1$.
Massless scalar field obeys the following equation (for its radial part):
\beq
\frac{\om^2 \phi(r)}{T(1-r)}+
\pr_r \l( T(1-r) \pr_r \phi(r) \r)  - p^2 \phi(r) = 0.
\eeq
Which has two solutions:
\beq
I_{\pm 2 i \omega/T} (2 p\sqrt{1-r}/\sqrt{T}),
\eeq
where $I$ is the Bessel function (\verb|BesselI| in Mathematica).
Since the imaginary part of $\omega$ should be negative (to ensure decay in time) and the in-going mode is growing, we pick the solution with $-2 i \omega/T$. To be definite, we impose mixed Robin boundary conditions at the screen:
\beq
a I_{-2 i \omega/T} (2 p\sqrt{1-r_*}/\sqrt{T}) +
b \pr_{r_*} I_{-2 i \omega/T} (2 p\sqrt{1-r_*}/\sqrt{T}) = 0.
\eeq
In the limit of small momenta $p$
we can Taylor expand the Bessel function to find the root. In this case the index $-2 i \omega/T$ has to be close to a negative integer. When it is close to minus one we get diffusive branch:
\beq
\omega = - i \l( \frac{T}{2} + D p^2 \r),
\eeq
where the diffusion coefficient depends on the location of the screen and the boundary condition, but not the temperature:
\beq
D = \frac{(r_*-1)(b+2a(r_*-1))}{2(b-2a(r_*-1))}.
\eeq
For Dirichlet boundary condition $b=0$, it simplifies to $D=(1-r_*)/2$. 
Interestingly, for Neumann case $a=0$ it is negative: $D=-(1-r_*)/2$. It does not lead to any pathologies because there is the leading $-iT/2$ which insures the decay. We call this behavior diffusive because $p^2$ dependence tells us how fast the excitations propagate.
When $-2i \omega/T$ is minus two or lower, we get sub-diffusive branches with $i\omega \sim p^4, p^6, \dots$.

\subsection{2+1 de Sitter--black hole}
\label{app:ds_bh}
The metric
\beq
ds_{BH}^2 = -(1-2M -r^2)dt^2 + \frac{dr^2}{1-2M-r^2}+r^2 d\varphi^2
\eeq
does not have a horizon at $r=0$. Instead, it is a conical singularity because $\varphi$ still have period $2\pi$. In fact, it is just a quotient of empty de Sitter. The wave equation can be written
as
\beq
\frac{\om^2}{1-2M-r^2} \phi + 
\frac{1}{r} \pr_r  \l( r (1-2M-r^2) \pr_r \phi \r) - \frac{l^2}{r^2} \phi - m^2 \phi=0,
\eeq
with $l$ being integer, since the angular part is $e^{il \varphi}$. One can get rid of $M$ by rescaling the variables:
\beq
\widetilde{r}= \frac{r}{\sqrt{1-2M}},
\eeq
\beq
\widetilde{\om}= \frac{\om}{\sqrt{1-2M}},
\eeq
\beq
\widetilde{l}= \frac{l}{\sqrt{1-2M}}.
\eeq
In this variables we have the same wave equation as in empty de Sitter. This change also does not affect the boundary conditions. Hence the (low) branch of quasinormal modes is (c.f. eq. (\ref{eq:low_branch})):
\beq
\omega = -i \l( l + \sqrt{1-2M}(\Delta+2k) \r), \ k=0,1,2,\dots.
\eeq
Linear $l-$dependence remains intact.

\subsection{Toy geometries}
\label{app:toy}
Let us start from geometry A:
\beq
ds_A^2 = -(1-r)dt^2 + \frac{dr^2}{1-r} + r^2 d\varphi^2.
\eeq
Massless wave equation admits an explicit solution. The one which is regular at $r=0$ is
\beq
\phi = (1-r)^{-i \omega} r^l \ 
\phantom{}_2 F_1 \l( \frac{1}{2}+l-i \om-\oh \sqrt{1-4\om^2}, \oh + l  - i \om+ \oh \sqrt{1-4 \om^2}, 1+2l,r \r).
\eeq
Performing the $r \ra 1-r$ transformation (\ref{eq:z_transform}), we require that $(1-r)^{+i \om}$ is absent because it corresponds to out-going mode at the horizon. This would be true if either of $\frac{1}{2}+l-i \om \pm \oh \sqrt{1-4\om^2}$ is a non-positive integer. Hence the quasi-normal modes are:
\beq
\omega = -i \frac{(k+l)(1+k+l)}{1+2k+2l}, \ k = 0,1,2,\dots.
\eeq
As we explained in the main text, localized perturbations correspond to large $l$. In this case we again see a linear $l-$dependence.

Finally, for the geometry B
\beq
ds_B^2 = -(1-r^2)dt^2 + \frac{dr^2}{1-r^2} + d\varphi^2.
\eeq
the wave equation again admits an explicit solution in terms of Legendre functions:
\beq
P_{(-1+\sqrt{1-4l^2})/2}^{i \omega}(r), \ 
Q_{(-1+\sqrt{1-4l^2})/2}^{i \omega}(r)
\eeq
Both of them are regular at $r=0$. Notice that now the spacetime does not cap-off at $r=0$. We can imagine we have some complicated boundary condition there which we can satisfy only if we have both modes. 
Legendre $P$ is purely out-going at the horizon, since it can be expressed as
\beq
P_{\la}^{i \om}(r) = \frac{1}{\Gamma(1-i \om)} 
\l( \frac{1+r}{1-r} \r)^{i\om/2} \ \phantom{}_2 F_1\l( -\la, \la+1,1-i \om, \frac{1-r}{2}\r), \ \ \la = \oh(-1+\sqrt{1-4l^2}),
\eeq
and the hypergeometric function has a regular expansion at $r=1$. 
The $Q$ function is a bit more complicated:
\beq
\begin{split}
    Q_{\la}^{i \om}(r) = \sqrt{\pi } &\frac{\Gamma(\la+i \om+1)}{2^{\la+1} \Gamma(\la+3/2)} e^{-\om \pi} r^{-\la-i\om-1} (1-r^2)^{i \om/2} \ 
\\
&\times \phantom{}_2 F_1 \l( \frac{\la+i \om+1}{2}, \frac{\la+i\om+2}{2},\la+3/2,1/r^2 \r).
\end{split}
\eeq
Again, performing (\ref{eq:z_transform}) we can enforce $Q$ being out-going as well 
if $(2+\la-i\om)/2$ or $(1+\la-i\om)/2$ is a non-positive integer. This leads to two family of frequencies:
\beq
\om_1 = \oh \sqrt{4l^2-1} -\frac{i}{2} \l(1+4k \r),
\eeq
and
\beq
\om_2 = \oh \sqrt{4l^2-1} -\frac{i}{2} \l(3+4k \r), \ k = 0,1,2,\dots. 
\eeq
For large $l$ they become sound modes which propagate ballistically ($\Re \omega \sim l$) with a constant damping.

\section{Bounds on OTOC}
\label{app:otocbound}
In this Appendix we give a proof on a lower bound at infinite temperature on OTOC by the local two-point functions.
The bound (\ref{eq: bound-otoc}) holds for any system, integrable or chaotic, local or non-local.
Bound (\ref{eq:bound_comp}) is a conjecture which should be true for chaotic systems at early times. We prove under a certain assumption about the operator-size distribution.

\subsection{Spin operators}
\label{app:spinOTOC}
We start with the time evolution of a Majorana operator at site $i$:
\beq
\psi_{i}\left(t\right)=e^{-iHt}\psi_{i}e^{iHt}
\eeq
where $H$ is some local Hamiltonian to be specified in the following discussion.  
We assume that $\psi_i$ obey the following anti-commutaton relations:
\beq
\{\psi_i, \psi_j\} = 2 \delta_{ij},
\eeq
but the subsequent can be easily generalized to a product of Majorana operators (including Pauli spin operators).
We can then introduce the 2 point function and out-of-time ordered correlation function as
\beq \label{eq:G-and-OTOC}
G(t) = \bra \psi_i(t) \psi_i (0)\ket \equiv \Tr (\psi_i(t) \psi_i (0)),\quad \text{OTOC}(t) = \bra \psi_i(t) \psi_j(0) \psi_i (t) \psi_j (0) .
\eeq
In this Section we normalize the trace so that the trace of identity is one.
We then establish the following inequality that bounds the $-$OTOC$(t)$ from below by $G(t)$ as:
\beq  \label{eq: bound-otoc}
-\text{OTOC}(t) \geq 2 G(t)^2 -1.
\eeq
To furnish the proof of \eqref{eq: bound-otoc} we introduce a local operator basis $\{\Gamma_I \}$, and formulate \eqref{eq: bound-otoc} as an inequality among the overlap coefficients between $\psi_i (t)$  and $\Gamma_I$.  One can always expand $\psi_i (t)$ as:
\beq \label{eq:expansion}
\psi_{i}\left(t\right)=\sum_{I}c_{I}\left(t\right)\Gamma_{I}.
\eeq
The time dependence is then encoded in the coefficients $c_I (t)$.  $\Gamma_I$ can be expanded as a product of spin operators as:
\beq 
 \Gamma_{I}=i^{k\left(k-1\right)/2}\prod_{j=1}^{|I|} \psi_{i_{j}},\quad I=(i_1,i_2,\dots,i_k)
\eeq
where $k=|I|$ is the length of the index set $I$.  The factor $i^{k\left(k-1\right)/2}$ is chosen such that 
\beq \label{eq: basis-norm}
\Gamma_{I}=\Gamma_{I}^{\da},\quad\tr\left(\Gamma_{I}\Gamma_{J}\right)=\del_{IJ}.
\eeq
They obey the following commutation relations:
\beq
\Gamma_{I^{\pp}}\Gamma_{I}=\left(-1\right)^{|I\cap I^{\pp}| + |I^\pp| |I|}\Gamma_{I}\Gamma_{I^{\pp}}
.
\eeq\label{eq:commutator-Gamma}
With this convenient choice of basis we can extract coefficient $c_I$ by taking trace with the corresponding basis, explicitly, we have
\beq
c_I (t) = \bra \Gamma_I | \psi_i (t) \ket = \tr(\Gamma_I \psi_i(t))
\eeq
Now we specify the interaction of the Hamiltonian as:
\beq \label{eq:Hint}
H=\sum_{I}J_{I}\Gamma_{I},\quad |I|\text{ even}
\eeq
We restrict to the case where all interacting terms in $H_{\text{int}}$  are of even length because a Hamiltonian is supposed to be bosonic. Similarly, time-evolved $\psi_1(t)$ involves only fermionic operators: $|I|$ in eq. (\ref{eq:expansion}) is odd.
 
 We move on to the two point function $G(t)$ and $\text{OTOC}(t)$. The two point function can be simply obtained as:
\beq
G\left(t\right)=\bra\psi_{i}\left(t\right)\psi_{i}\left(0\right)\ket=\sum_{J}c_{J}\tr\Gamma_{J}\psi_{i}=c_{i}\left(t\right),
\eeq
which can be interpreted as the part of $\psi_i (t)$ that localizes at its original position $i$. Because $\psi_i (t)^2=1$, $c_I^2$ define a probability distribution:
\beq
\label{eq:cnorm}
\psi_i(t)^2 = 1 = \sum_I c_I^2.
\eeq
Now let's consider the expression of OTOC. We expand the two $\psi_i (t)$s in \eqref{eq:G-and-OTOC} into operator basis as: 
\beq \label{eq:OTOC-expansion}
-\text{OTOC}(t)=- \sum_{J_{1},J_{2}}c_{J_{1}}c_{J_{2}}\tr\left(\Gamma_{J_{1}}\psi_{j}\Gamma_{J_{2}}\psi_{j}\right)
\eeq
The commutation relation between $\psi_j$ and $\Gamma_J$ is given by:
\beq
\psi_{j}\Gamma_{J}  
 =\begin{cases}
\Gamma_{J}\psi_{j} & j\in J\\
-\Gamma_{J}\psi_{j} & j\not\in J
\end{cases}
\eeq
where in the above derivation, we have assumed that $|J|$ is odd, which applies to the case of $J_1$ and $J_2$ in \eqref{eq:OTOC-expansion}. We can then commute the first $\psi_j$ with $\Gamma_{J_2}$ in the equation. This splits the sum over $J_2$ into two pieces: for $j\not\in J_2$, the commutation leads to an extra minus sign, while for $j\in J_2$, it does not.  The equation \eqref{eq:OTOC-expansion} can be recast as:
\beq\begin{split}
- \text{OTOC}(t) &=-\sum_{J_{1}}\sum_{j\in J_{2}}c_{J_{1}}c_{J_{2}}\tr\left(\Gamma_{J_{1}}\Gamma_{J_{2}}\right)+\sum_{J_{1}}\sum_{j\not\in J_{2}}c_{J_{1}}c_{J_{2}}\tr\left(\Gamma_{J_{1}}\Gamma_{J_{2}}\right)\\
 & =-\sum_{j\in J_{2}} c_{J_{2}}^{2}+\sum_{j\not\in J_{2}}c_{J_{2}}^{2}.
\end{split}\eeq
In deriving the second equality, we have used the condition $\tr\Gamma_{J_1}\Gamma_{J_2} = \delta_{J_1 J_2}$ and the sum over $J_1$ leads amounts to setting $c_{J_1}$ to be equal to $c_{J_2}$. Now we can use eq. (\ref{eq:cnorm}) to convert $j\in J_2$ terms above as:
\beq
\sum_{j\in J_2} c^{2}_{J_2} = 1 - \sum_{j\not\in J_2} c^{2}_{J_2}.
\eeq
Therefore, we arrive at the following expression of $-$OTOC$(t)$ as:
\beq
-\text{OTOC}(t)= 2 \sum_{j\not\in J_{2}} c_{J_{2}}^{2}-1.
\eeq
In the case where $j\not=i$, we know $c^{2}_i$ appears as one term in the sum of the right hand side, and therefore, we conclude that 
\beq
-\text{OTOC}(t)= 2 c^{2}_{i} (t) -1 \geq 2 G^2 (t) -1.
\eeq
This proves \eqref{eq: bound-otoc} for the case $i\not=j$.  
One comment: the derivation is definitely true for a fixed Hamiltonian. Under the possible disorder average:
\beq
-\bra \text{OTOC} \ket_J \ge 2 \bra G(t)^2 \ket_J - 1 \ge 2 \bra G(t) \ket^2_J - 1. 
\eeq

The above results extend to the case of averaged two point function and OTOC, defined as:
\beq \label{eq:OTOCA}
G\left(t\right)=\bra\psi_{i}\left(t\right)\psi_{i}(0)\ket,\quad \text{OTOC \ensuremath{\left(t\right)}}_A=\frac{1}{N}\sum_{j}\bra\psi_{i}\left(t\right)\psi_{j}(0)\psi_{i}\left(t\right)\psi_{j}(0)\ket.
\eeq 
 One can show \eqref{eq: bound-otoc} is still valid following the same strategy as above. In the following we present an alternative argument based on the property of operator size.
Note that the sum in the definition of OTOC$_A$ includes $j=1$ term, but this term is always less than $1/N$ so it will not be able invalidate the bound at large $N$.

The OTOC defined in \eqref{eq:OTOCA} is equal to the operator size $S$:
\beq
\label{eq:otoc-size}
-\OTOC_A = 1 - \frac{2 S(t)}{N},
\eeq
\beq
S(t) = \sum_I |I| |c_I|^2.
\eeq
As we know, $c_1 = G$, so the rest of $c_I^2$ sum to $1-G^2$. But the maximum size is $N$. It means that the OTOC is bounded by
\beq
-\OTOC_A \ge 1 - \frac{2}{N}(G^2 + N(1-G^2) ) \gtrsim 2G^2 - 1.
\eeq
Which is exactly the bound we derived earlier.

This bound is, in fact, saturated. Consider 2-Majorana (1-qubit) Hamiltonian $H = i\psi_1 \psi_2 = \sigma_z$.
We can compute OTOC and $G$ explicitly, by using the fact that $H$ anti-commutes with everything:
\beq
G = \oh \Tr (e^{-2 i H t}) = \cos(2t),
\eeq
\beq
-\text{OTOC} = \oh \Tr e^{-4 i H t} = \cos(4t) = 2 \cos^2 (2t) - 1 = 2G^2-1.
\eeq
More generally, this bound will be saturated for the Hamiltonian $H=\psi_1 \psi_2 \dots \psi_N$.
This Hamiltonian is unphysical because it mixes all fermions at the same time. This results in a highly atypical operator size distribution:
$\psi_1(t)$ has support only on $\psi_1$ and $\psi_2 \psi_3 \dots \psi_N$.

Can we use any extra assumptions to derive a stronger bound?
For simplicity, let us consider the system with no spacial structure so that we can derive everything from the size distribution.
Naive expectation is that for generic systems the size distribution $P_s(t)$  defined as
\beq
P_s(t) = \sum_{I: |I|=s} |c_I|^2,
\eeq
should monotonically decrease with $s$: at $t=0$ the distribution is a delta-function peak at $s=1$ which then relaxes. This is indeed true for the SYK model \cite{Roberts:2018mnp} at early times. At least it should be true starting from some size $\Lambda$. For example, if a Hamiltonian has a term which mixes $4$ fermions and another term mixing $100$ of them then for small operator sizes there could be some non-monotonicity in the operator distribution due to a peak at $s=100$. But for larger sizes we can expect that the distribution is monotonic. 
Hence, our main assumption is:

\begin{figure}
    \centering
    \includegraphics[scale=1.2]{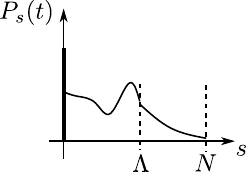}
    \caption{The size distribution for which we prove eq. (\ref{eq:bound_comp}).
    There is a delta-function at $s=1$, then some distribution for $1<s<\Lambda$ for which we remain agnostic and after $s>\Lambda$ it is monotonically decreasing.}
    \label{fig:distribution}
\end{figure}

\textbf{Assumption:} the probability distribution defined by $|c_I|^2$ is monotonically decreasing for $s \ge \Lambda = \epsilon N, \epsilon<1/2$. A characteristic $P_s(t)$ is depicted on Figure \ref{fig:distribution}.

\textbf{Claim:} in this case
\beq
\label{eq:bound_comp}
-\OTOC_A \ge G^2  - \epsilon (1-G^2) - \frac{2}{N}G^2.
\eeq

\textbf{Proof:} to bound the averaged OTOC (\ref{eq:otoc-size}) from below we need to maximize the expectation value of the size $S$. It will be maximized is the distribution is constant on the interval $\Lambda < s <N$. Suppose this constant is $v$.
Since there is a delta-function peak at $s=1$ with the weight at least $G^2$,
away from $s=1$ the probability is bounded by $1-G^2$. Hence:
\beq
\text{Pr}(0 < s < \Lambda) + v(N-\Lambda) \le 1-G^2.
\eeq
The expectation value $S$ is
\begin{align}
S = \int_0^N ds \ s P_s(t) \le 
1 \times G^2 + \Lambda \times \text{Pr}(0 < s < \Lambda) + \frac{(N^2-\Lambda^2)}{2} v \le \\ \nonumber
\le G^2 + \frac{N+\Lambda}{2} (\text{Pr}(0 < s < \Lambda) + (N-\Lambda) v) \le G^2 + \frac{N+\epsilon N}{2}(1-G^2).
\end{align}
In the second inequality we used the fact that $\Lambda < N/2$. The last inequality is what we wanted to prove. 
If $\epsilon$ is small, then in the large $N$ limit we expect something like
\beq
-\OTOC_A \gtrsim G^2.
\eeq
In the next Subsections we corroborate this by studying "typical" operators and demonstrating this in simple chaotic systems.

\subsection{Typical operators}
\label{app:typical}
\begin{figure}
    \centering
    \includegraphics{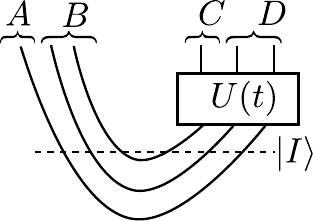}
    \caption{The illustration of how
    the maximally entangled state is prepared and how we split the system after acting with $U$.}
    \label{fig:Istate}
\end{figure}

Ref. \cite{Hosur:2015ylk} introduced a useful framework for quantifying scrambling properties
of unitaries. It will allow us to make statements about OTOCs and two-point functions of \textit{generic} operators. It is also ideal for quantifying scrambling if one asks about the maximally mixed initial state. 
We double the system and  prepare a maximally entangled state $|I\ket$ which has two sides - Figure \ref{fig:Istate}. After that we act with the unitary of interest on one side and ask about the correlations between different parts, say $AB$ and $CD$. The split of one side into $AB$ (or $CD$) can be arbitrary. 

Initially, in state $| I \ket$ the two sides are perfectly correlated.
After we act with $U$ we expect that the correlation is highly diminished. 
A way to quantify it is to compute the mutual information $I(A:C)$. It turns out it is directly related to the averaged OTOC \cite{Hosur:2015ylk, Lensky:2018hwa}:
\beq
\mathbb{E}_{AD} \bra AD A D \ket = 4^{-a} e^{I_2(A:C,t)}
\eeq
where we used the concise notation $\bra A D A D \ket$ for the correlation function
$ \bra I  |  \Oc_A U^\dagger \Oc_D U \Oc_A U^\dagger \Oc_D U | I \ket $ and the expectation value $\ee_{AD}$ means the average over a complete basis of operators in $A$ and $D$, there are $4^{a+d}$ of them.
$I_2(A:C)$ is the second Renyi mutual information between $A$ and $C$.

Such expectation is not very interesting because it contains the identity operators in $\Oc_{A,D}$,
for which the OTOC does not decay. The contribution of the identity operators is
$4^a+4^d-1$.
 Subtracting them
produces an expectation value which we denote as $\bra \dots \ket_{int}$:
\beq
\mathbb{E}_{AD} \bra AD A D \ket_{int} = 4^{-a} \l( e^{I_2(A:C)} - 1 \r) - 4^{-d} + 4^{-a-d},
\eeq

Similarly, one can average \cite{Lensky:2018hwa} the square of the two-point function $\bra I | \Oc_A \Oc_C | I \ket$.
Again, subtracting the identity operator we get
\beq
\ee_{AC} \bra A C \ket_{int}^2 = 4^{-a-c} \l( e^{I_2(A:C,t)} -1 \r)
\eeq
We obtain the following equality for any evolution operator $U(t)$:
\beq
\mathbb{E}_{AD} \bra AD A D \ket_{int} = 4^c \ee_{AC} \bra A C \ket_{int}^2 - 4^{-d} + 4^{-a-d}.
\eeq
Usually we compute OTOC with local operators, so $D$ should be seen as a small subsystem.
Hence its complement $C$ is huge. We can further restrict on a subset $C_1$ of $C$. This way
we get an inequality instead of inequality:
\beq
\mathbb{E}_{AD} \bra AD A D \ket_{int} \ge 4^{c_1} \ee_{AC_1} \bra A C_1 \ket_{int}^2 - 4^{-d} + 4^{-a-d}.
\eeq

If $A,D$ are small subsystems then it is hard to infer anything because $4^{-d}, 4^{-a-d}$ can be important. In this case one can consult Appendix \ref{app:spinOTOC} about the spin operators.
Instead, let consider the situation when $A,D,C_1$ contain a lot of degrees of freedom, but not necessarily a macroscopic fraction of the whole system.
Assume subsystems $A,C$ each has $a$ spins and all \textit{diagonal} correlators (the ones having the same operator in $A$ and $C$) involving 
more than $l$ spins are large, say larger than a certain number $g$.
This models the situation with the double-scaled SYK, where
the correlators which are under the analytical control are the ones including a lot of spins.
As long as $1 \ll l \ll a$ it is enough to ensure that the average correlator is larger than $g$.
The fraction of operators supported on more than $l$ spins is:
\beq
4^{-a}\sum_{x=l}^a \bin{a}{x} 3^x = 1 - \sum_{x=0}^{l-1} \bin{a}{x} 3^x \approx
1 - 4^{-a} \sum_{x=0}^{l-1} \frac{a^x}{x!} 3^x = 1 -  4^{-a} e^{3a} \frac{\Gamma(l,3a)}{\Gamma(a)} 
\approx 1 - 4^{-a} a^l \frac{3^{l}}{3a}.
\eeq
For both approximations we used $a \gg l$.
The last expression goes to 1 for $a \gg l$. So up to corrections which go as $4^{-d}, 4^{-a}$ we get a bound
\beq
\label{eq:averbound}
\mathbb{E}_{AD} \bra AD A D \ket_{int} \ge g^2.
\eeq
The factor $4^{c_1}$ disappeared because we considered the diagonal correlators.
To summarize, if a large fraction of two-point functions is greater than $g$, it implies that there exists OTOC with a value greater than $g^2$.


\subsection{Some numerical results}
\label{app:otocnum}
We can study two-point function and OTOC numerically in various finite systems. For convenience, let us introduce two place-holder operators $A$ and $D$, so that the quantities we look at are:
\beq
G = \Tr( A(t) A(0) ),
\eeq
\beq
\OTOC = \Tr( A(0) D(t) A(0) D(t) ) .
\eeq

Our expectation is that 
\beq
|\OTOC| \ge G^2,
\eeq
at least before $G$ becomes zero. It corresponds to the time interval when the Heisenberg-evolved operator is still has support on the initial location, hence we can expect the operator-size distribution to have a decaying tail.

We will consider SYK and chaotic Ising-like spin chain with the Hamiltonian
\beq
\label{eq:spin_chain_H}
H_{chaotic} = \sum_{i=1}^N Z_i Z_{i+1} - 1.05 X_i + 0.5 Z_i + 0.1 Y_i Z_{i+1},
\eeq
where we are assuming periodic boundary conditions.
We will contrast our results with the integrable XX chain:
\beq
\label{eq:XX_chain}
H_{XX} = \sum_{i=1}^N X_i X_{i+1} + Y_i Y_{i+1},
\eeq
again, with periodic boundary conditions. This model has a next-to-nearest neighbor (NNNI) deformation which preserves integrability in the first order in perturbation theory \cite{Jung_2006}:
\beq
\label{eq:xx_def}
H_{XX+NNNI} = H_{XX} + g \sum_{i=1}^N Z_i Z_{i+2}.
\eeq

\paragraph{Local operators.}

\begin{figure}
    \centering
    \minipage{0.5\textwidth}
    \centering
    \includegraphics[scale=1.]{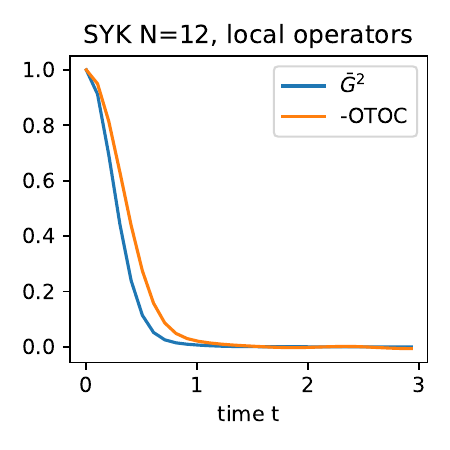}
    \endminipage
    \minipage{0.5\textwidth}
    \centering
    \includegraphics[scale=1.]{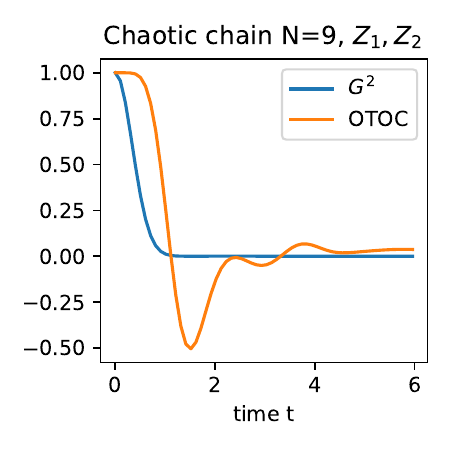}
    \endminipage

\minipage{0.5\textwidth}
    \centering \includegraphics[scale=1.]{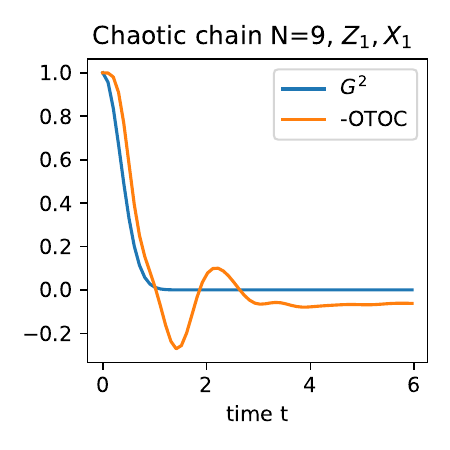}
    \endminipage
\minipage{0.5\textwidth}
    \centering \includegraphics[scale=1.]{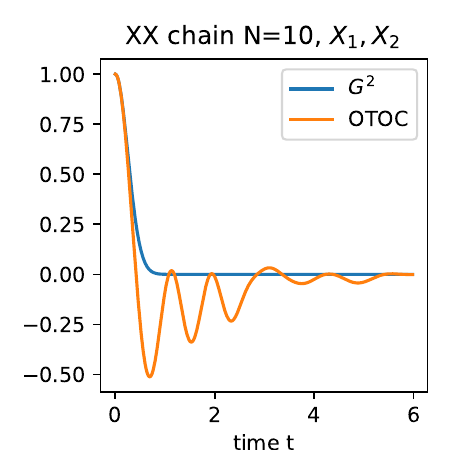}
    \endminipage

    \minipage{0.5\textwidth}
    \centering \includegraphics[scale=1.]{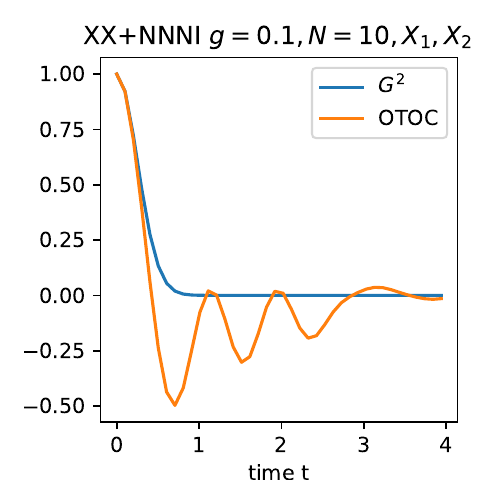}
    \endminipage
\minipage{0.5\textwidth}
    \centering \includegraphics[scale=1.]{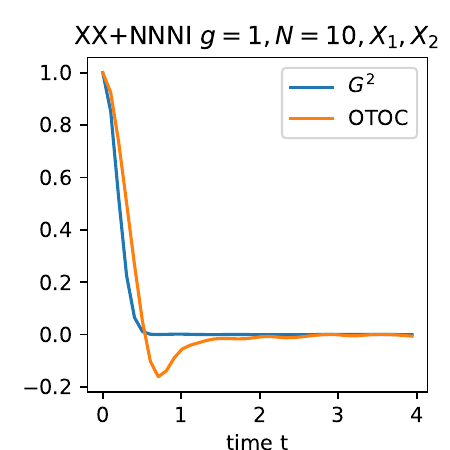}
    \endminipage
    
    \caption{Decay of two-point function squared (blue) vs the decay of OTOC (orange) for various local operators and systems.}
    \label{fig:num_local}
\end{figure}

In SYK, we study the standard two-point function and OTOC, so the operators $A,D$ are:
\beq
\text{SYK:} \ A=\psi_1, \ D=\psi_2.
\eeq
In the spin chain we study local magnetization
\beq
\text{spin chain:} \ A=Z_1, \ D=Z_2 \ \text{or} \ D=X_1.
\eeq
The results are presented on Figure \ref{fig:num_local}. One can expect that OTOC does not decay initially because of the Lieb--Robinson bound, but for the adjacent sites this effect is minimal. In fact, $G$ and OTOC reach zero approximately at the same time.
The bound $\text{OTOC} \ge G^2$ is violated for the integrable XX chain and the near-integrable XX chain, when the perturbation parameter $g$ is small (e.g. $g=0.1$). However, sufficiently far away from the integrable point ($g \gtrsim 1$), this inequality is satisfied.

\paragraph{Non-local operators.}

\begin{figure}
    \centering

\minipage{0.5\textwidth}
    \centering
\includegraphics{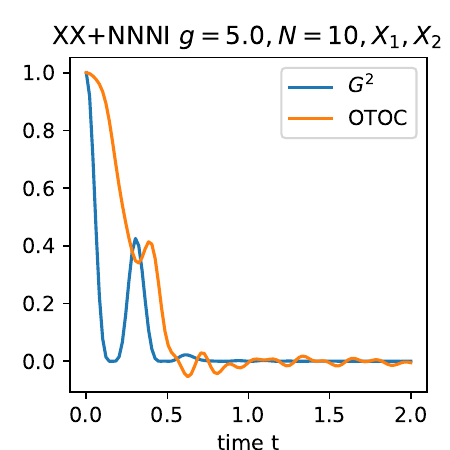}
\endminipage
\minipage{0.5\textwidth}
    \centering
\includegraphics{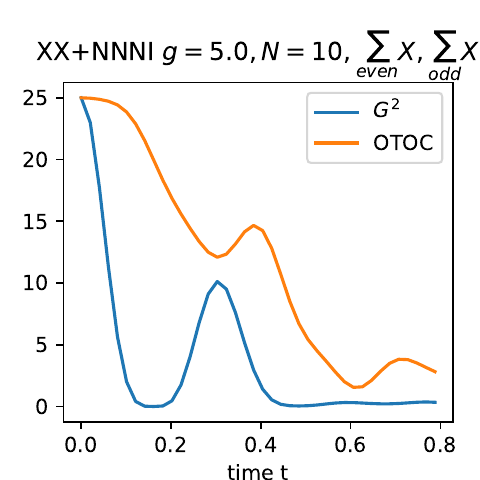}
\endminipage
    
    \caption{Correlation functions in the deformed XX chain (\ref{eq:xx_def}) when the integrability breaking parameter is large.}
    \label{fig:large_g}
\end{figure}

Let us consider the following two operators in the spin chain (\ref{eq:spin_chain_H}):
\beq
A = \sum_{odd} Z_i,
\eeq
\beq
D = \sum_{even} Z_i.
\eeq
\begin{figure}
   
    
    \minipage{0.5\textwidth}
    \centering
\includegraphics{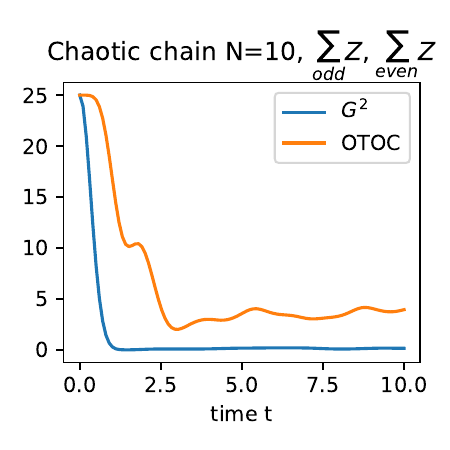}
\endminipage
 \minipage{0.5\textwidth}
    \centering
\includegraphics{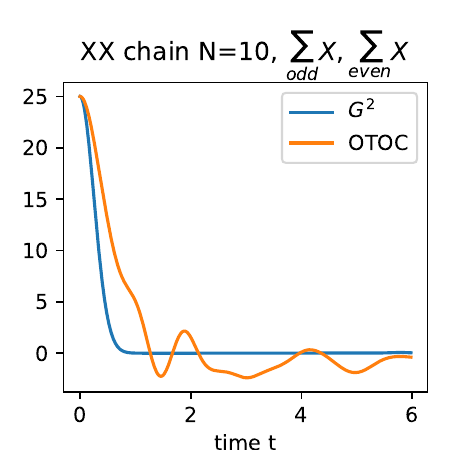}
\endminipage

 \minipage{0.5\textwidth}
    \centering
\includegraphics{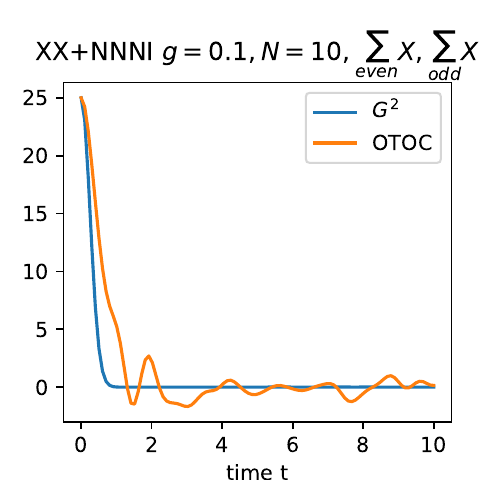}
\endminipage
 \minipage{0.5\textwidth}
    \centering
\includegraphics{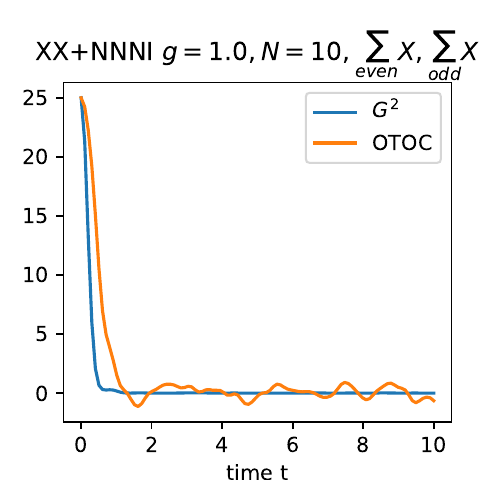}
\endminipage

    \caption{Non-local correlators in various physical systems.}
    
    \label{fig:nonlocal}
\end{figure}
Interestingly, we find that $|\text{OTOC}| \ge G^2$ is satisfied even for integrable XX model - Figure \ref{fig:nonlocal}.
Finally, we also show the behavior of correlators for the perturbed XX chain (\ref{eq:xx_def}) when the chaotic deformation parameter is large - Figure \ref{fig:large_g}. The behavior of correlation functions is qualitatively different, but the bound is still obeyed, at least before $G$ becomes small for the first time.

The behavior of OTOC with non-local operators was recently addressed in \cite{Zhou:2021syv} and it was argued that their behavior is similar to the behavior of local operator OTOCs.
For concreteness, one can look at the total spin operator
\beq
Z = \sum_i Z_i,
\eeq
and the corresponding OTOC:
\beq
\label{eq:globalOTOC}
\Tr([Z(t),Z]^2) = 
\sum_{ij} \Tr([Z_i(t),Z_j]^2) + \sum_{i \neq k \ \text{or} \ j \neq l}
\Tr([Z_i(t),Z_j][Z_k(t), Z_l]).
\eeq
The first sum is the sum of "normal" OTOCs.
Ref. \cite{Zhou:2021syv} argued
that the second, "off-diagonal", contribution is small.
The Heisenberg evolution of $Z_i(t)$ contains a sum over different operators $\Oc_x$:
\beq
\label{eq:HeisenbergO}
Z_i(t) = \sum_x c_x \Oc_x.
\eeq
Commutator $[Z_i(t),Z_j]$
picks operators in $Z_i(t)$ which do not commute with $Z_j$, and similarly for $[Z_k(t), Z_l]$. Taking a product and a trace computes the overlap of such operators. For generic $ijkl$ such overlap is small. Moreover, the corresponding product of the expansion coefficients $a_{x} a_{x'}$ will have an oscillatory phase leading to cancellations. In contrast, the first "diagonal" term in (\ref{eq:globalOTOC}) picks up the absolute value $a_x^2$.

\section{Gravitationally Dressed Stretched Horizon and Tomperature for Distinct Observers}
\label{app:dressing}
The comparison of decay rates  involves different states, it is then a natural question of how to implement such comparison in a putative gravitational theory that involves the static patch of de Sitter. We consider states that contain excitations above the vacuum, where the semi-classical description remains valid and then examine whether the vacuum state is the one that exhibits the fastest decay. One is then motivated to compare the “tomperature” for different states in this context. 

The above setup suffers from several issues. First of all, it is not clear what the gauge invariant bulk time that corresponds to the time in the dual theory is, and which observer to choose when the measurement is conduct. This follows from the essential difference between static patch physics and holographic AdS/CFT: while in latter case the observer is unambiguously defined at boundary region with physical Hamiltonian, in the former case the observer is part of the system and there is no explicit physical Hamiltonian that generates physical time evolution. Therefore, we examine in the subsequent discussion the tomperature for the observers living both at the stretched horizon and the pole, and claim that the tomperature exhibits different behavior against perturbations around the static patch vacuum. 

The second issues involves the definition of stretched horizon. It's defined as a codimension-1 surface which is at a few Planck's lengths away from the static patch horizon. Note that the size of static patch dynamically changes with perturbations to the static patch vacuum, therefore, one should incorporate those dynamical aspects to the definition of the stretched horizon. In the following discussion we propose a fix by gravitationally dressing the stretched horizon to the thermodynamical volume of static patch in the following discussion. 

Finally, we examine the tomperature for distinct observers in Schwarzschild de-Sitter with small black hole mass. We intend to view this as a small perturbation to the static patch vacuum. We show that for stretched horizon observer, the tomperature monotonically decreases for increasing small mass of the black hole, while for pole observer it increases. We provide further speculations on this fact at the end of this Section.

\subsection{Stretched Horizon Dressed to the Thermodynamic Volume of de-Sitter}
In this Section we review the definition of stretched horizon in static
patch vacuum and propose its gravitational dressing to thermodynamical
volume of de Sitter. In vacuum static patch, the stretched horizon
is defined in global coordinates such that its a few Planck distance
away from the horizon $r=r_{h}$. 
\beq
r_{sh}=R_{dS}-\#G_N=\alpha_{0} R_{dS}
\eeq
where $R_{dS}$ is the de-Sitter radius, and in the second equality, we
defined the ratio between $r_{sh}$ and $R_{dS}$ to be $\alpha$, which
is a constant and only depends on the Newton's constant. Now let's
consider turning on some small mass $M$ in the static patch which
forms a black hole. This changes the size of static patch. In \cite{Dolan_2013}, the authors defined thermodynamical volume $V_{{\rm dS}}$
that characterizes the dynamical volume of the static patch. Using the static path metric
\beq
\dd s^{2}=-f\left(r\right)\dd t^{2}+f(r)^{-1}\dd r^{2}+r^{2}\dd\Omega_{d-2}^{2},\quad f\left(r\right)=1-\frac{2M}{r^{d-3}}-\frac{r^{2}}{R_{dS}^{2}},
\eeq
the thermodynamical volume of Schwarzschild black hole
is defined as: 
\beq
V_{{\rm dS}}=\frac{{\rm Vol}\left(S^{d-2}\right)}{d-1}\left(r_{h}^{d-1}-r_{b}^{d-1}\right).
\eeq
where $r_{h}$ is the cosmological horizon radius and $r_{b}$ is
the black hole horizon radius. They depends on the mass of the black
hole and therefore, $V_{{\rm dS}}$ is a function on the black hole
mass. We then define the dynamical ratio $\alpha\left(M\right)$ by
setting:
\beq \label{eq:ratio}
\frac{\alpha\left(M\right)}{\alpha\left(M=0\right)}=\left(\frac{V_{{\rm dS}}\left(M\right)}{V_{{\rm dS}}\left(0\right)}\right)^{\frac{1}{d-1}}
\eeq
where $\alpha_{0}=\alpha\left(M=0\right)$ is the ratio we defined
in static patch vacuum. It scales with the dynamical scale of de Sitter,
and we can therefore extend the definition of stretched horizon to
the Schwarzschild black hole case, with $r_{sh}\left(M\right)=\alpha\left(M\right)r_{h}\left(M\right)$.
Note that the thermal dynamical volume of vacuum static patch reads:
\beq
V_{{\rm dS}}\left(0\right)=\frac{{\rm Vol}\left(S^{d-2}\right)}{d-1}r_{h}\left(0\right)^{d-1}=\frac{{\rm Vol}\left(S^{d-1}\right)}{d-1} R_{dS}^{d-1}.
\eeq
Therefore we can explicit evaluate $\alpha\left(M\right)$ as: 
\beq
\alpha\left(M\right)=\alpha_{0}\left(\left(\frac{r_{h}\left(M\right)}{r_{h}\left(0\right)}\right)^{d-1}-\left(\frac{r_{b}\left(M\right)}{r_{h}\left(0\right)}\right)^{d-1}\right)^{\frac{1}{d-1}}
\eeq

\subsection{Stretched Horizon Observer}
Now let's calculate the tomperature for an observer at gravitationally-dressed
stretched horizon. We evaluate
the tomperature in the regime where $M/R_{dS}^{d-3}\ll1$. In this regime,
the horizon radius can be solved as:
\beq 
\begin{split}
    r_{h}\left(M\right) & =R_{dS}\left(1-\frac{1}{2}\left(\frac{2M}{R_{dS}^{d-3}}\right)+O\left(\frac{M}{R_{dS}^{d-3}}\right)^{2}\right),\\
r_{b}\left(M\right) & =R_{dS}\left(\left(\frac{2M}{R_{dS}^{d-3}}\right)^{1/(d-3)}+O\left(\left(\frac{M}{R_{dS}^{d-3}}\right)^{3/(d-3)}\right)\right),
\end{split}
\eeq 
where we keep our results up to the lowest order in $M/l^{d-3}$. Note
that the result for the black hole horizon radius $r_{b}\left(M\right)$
is only valud for $d\geq4$, namely, for bulk dimensions no less than
$4$. This makes sense because the lower dimensional de-Sitter black
holes do not form a horizon. The $\left(M/R_{dS}^{d-3}\right)^{1/(d-3)}$ dependence
is noteworthy, as for $d\geq5$ it dominates over any positive integer
powers. We will show its significant role in the mass dependence in
tomperature later. 
We can then move on to the evaluation of the surface gravity, which
to leading order is given by 
\beq
\kappa_{h}\left(M\right)=R_{dS}^{-1}\left(1-\frac{1}{2}\left(\left(d-2\right)\right)\frac{2M}{R_{dS}^{d-3}}+O\left(\frac{M}{R_{dS}^{d-3}}\right)^{2}\right)
\label{eq:kappa}
\eeq
which decreases with $M$ for sufficiently small $M$. Finally, we
compute the ratio factor via \eqref{eq:ratio} as:
\beq
\frac{\alpha\left(M\right)}{\alpha_{0}}=1-\frac{1}{2}\frac{2M}{R_{dS}^{d-3}}-\frac{1}{d-1}\left(\frac{2M}{R_{dS}^{d-3}}\right)^{\frac{d-1}{d-3}}+\dots
\eeq
where we have kept leading terms of both integer and fractional branches.
The gravitationally dressed stretched horizon is then located at $r_{sh}=\alpha\left(M\right)r_{h}\left(M\right)$,
which is given by
\beq
r_{sh}\left(M\right)=\alpha_{0} R_{dS} \left(1-\frac{2M}{R_{dS}^{d-3}}-\frac{1}{d-1}\left(\frac{2M}{R_{dS}^{d-3}}\right)^{\frac{d-1}{d-3}}+\dots\right)
\eeq
Therefore, for an observer sitting at the stretched horizon, with
$r=r_{sh}$, the associated ``tomperature'' is: 
\beq
T_{sh}\left(M\right)=\frac{\ka_{h}\left(M\right)}{\sqrt{f\left(r_{sh}\left(M\right)\right)}}
\eeq
namely, it's measured with respect to the observer's proper time.
We find in this case, the result turns out to be:
\beq
T_{sh}\left(M\right)=\ka_{sh}^{\left(0\right)}\left(1-\left(d-2+\frac{1}{\left(1-\alpha_{0}^{2}\right)\alpha_{0}^{d-3}}\right)\frac{M}{R_{dS}^{d-3}}-\frac{\alpha_{0}^{2}}{d-1}\left(\frac{2M}{R_{dS}^{d-3}}\right)^{\frac{d-1}{d-3}}+\dots\right)
\eeq
where leading order corrections from both branch exhibit a minus sign.
Therefore, this indicates that for a stretched horizon observer, the
tomperature is maximally decaying in static patch vacuum, while any
perturbations to it slows down the decay rate by decreasing the tomperature. 
\subsection{Geodesic Observer}
In this subsection we derive the tomperature for a geodesic observer,
whose location is given by the condition
\beq
f^{\pp}\left(r_{p}\right)=0.
\eeq
This corresponds to a non-spinning time-like geodesic located at constant $r$.
It is a natural generalization of a pole observer for empty de Sitter.
The pole observer is then sitting at:
\beq
r_{p}=\left(M(d-3) R_{dS}^{2}\right)^{\frac{1}{d-1}}
\eeq
where we can write the lapse function as:
\beq
f\left(r_{p}\right)=1-c\left(d\right)\left(\frac{2M}{R_{dS}^{d-3}}\right)^{\frac{2}{d-1}},\quad c\left(d\right)=\left(\frac{d-3}{2}\right)^{\frac{2}{d-1}}+\left(\frac{2}{d-3}\right)^{\frac{d-3}{d-1}}
\eeq
Therefore, we evaluate the tomperature for pole observer as:
\beq
T_{p}\left(M\right)=\frac{\ka_{h}\left(M\right)}{\sqrt{f\left(r_{p}\right)}}=R_{dS}^{-1}\left(1+\frac{1}{2}c\left(d\right)\left(\frac{2M}{R_{dS}^{d-3}}\right)^{\frac{2}{d-1}}-\frac{d-2}{2}\left(\frac{2M}{R_{dS}^{d-3}}\right)+\dots\right)
\eeq
Note that for $d\geq4$, the non-integer term always dominates over
the term with integer exponent, and shows that $T_{p}\left(M\right)$
grows with $M$ for sufficiently small $M$. This indicates that for
a pole observer, the perturbation to static patch vacuum can speed
up the decay rate. 

\subsection{Observer dressed to the sphere area}
Finally, we can look at the family of observers which observe the same area of the transverse sphere in different spacetimes. This corresponds to fixing the $r=r_s$ coordinate.  
Using eq. (\ref{eq:kappa}) we get
\beq
T_s(M) = \frac{\kappa_h(M)}{\sqrt{f(r_s)}} = 
\frac{1}{R_{dS} \sqrt{1-r_s^2/R_{dS}^2}} + M \frac{R_{dS} r_s^3 - (d-2) r_s^{d} R_{dS}^{2-d}(R_{dS}^2-r_s^2)}{ r_s^{d} (R_{dS}^2-r_s^2)\sqrt{1-r_s^2/R_{dS}^2}} + \dots.
\eeq
The leading $M$ correction is always positive. Thus the observed decay rate will be the slowest in empty de Sitter.

\printbibliography
\end{document}